\documentclass[preprintnumbers,superscriptaddress,showpacs,pra]{revtex4}%
\usepackage[colorlinks=true,linkcolor=blue]{hyperref}
\usepackage{amssymb}
\usepackage{amsfonts}
\usepackage{amsmath}
\usepackage{graphicx}%
\providecommand{\U}[1]{\protect\rule{.1in}{.1in}}

\let\stdsection\section
\renewcommand\section{\nopagebreak\stdsection}
\begin{document}
\title{No existence of the geometric potential for a Dirac fermion on a
two-dimensional curved surface of revolution }
\author{Z. Q. Yang}
\affiliation{School for Theoretical Physics, School of Physics and Electronics, Hunan
University, Changsha 410082, China}
\author{X. Y. Zhou}
\affiliation{School for Theoretical Physics, School of Physics and Electronics, Hunan
University, Changsha 410082, China}
\author{Z. Li}
\affiliation{School for Theoretical Physics, School of Physics and Electronics, Hunan
University, Changsha 410082, China}
\author{W. K. Du}
\affiliation{School for Theoretical Physics, School of Physics and Electronics, Hunan
University, Changsha 410082, China}
\author{Q. H. Liu}
\email{quanhuiliu@gmail.com}
\affiliation{School for Theoretical Physics, School of Physics and Electronics, Hunan
University, Changsha 410082, China}
\affiliation{Synergetic Innovation Center for Quantum Effects and Applications (SICQEA),
Hunan Normal University, Changsha 410081, China}
\date{\today}

\begin{abstract}
For a free particle that non-relativistically moves on a curved surface, there
are curvature-induced quantum potentials that significantly influence the
surface quantum states, but the experimental results in topological
insulators, whenever curved or not, indicate no evidence of such a potential,
implying that there does not exist such a quantum potential for the
relativistic particles, constrained on the surface or not. Within the
framework of Dirac quantization scheme, we demonstrate a general result that
for a Dirac fermion on a two-dimensional curved surface of revolution, no
curvature-induced quantum potential is permissible.

\end{abstract}

\pacs{03.65.Pm Relativistic wave equations; 04.60.Ds Canonical quantization;
04.62.+v Quantum fields in curved spacetime; 98.80.Jk Mathematical and
relativistic aspects of cosmology}
\keywords{Relativistic wave equations; Curved surface; Canonical quantization; Momentum}\maketitle

\section{Introduction}

The discovery of topological insulators has initialized a new era of condensed
matter physics \cite{topo05,topo07,topo10,topo11,2010PR,LeeDH,Iorio}. However,
on one hand, the surface quantum states are experimentally observed in
two-dimensional curved or flat topological insulators in which no
curvature-induced quantum potential is included. On the other hand, it seems
well established that, once a particle is constrained to remain on a curved
surface, there may be such a curvature-induced potential, conveniently called
as the \textit{geometric potential }\cite{jk,dacosta,fc,exp1,exp2}.
Explicitly, for a non-relativistic particle on the curved surface, the
\textit{geometric potentials} have been theoretically and experimentally
explored \cite{jk,dacosta,fc,exp1,exp2}, respectively. By the surface quantum
states of the topological insulators, we mean the two-dimensional states for
relativistic spin $1/2$ particles, usually with zero mass, and is is the Dirac
fermions as commonly called. There may be no existence of the geometric
potential for a Dirac fermion on two-dimensional curved surfaces. The present
paper deals with this problem.

The main aim of the present study is to show that for a two-dimensional
surface of revolution, such as sphere \cite{LeeDH}, Beltrami pseudosphere
\cite{Iorio}, cylinder \cite{cylinder}, helicoid \cite{helicoid}, etc.
\cite{book} Dirac quantization scheme offers a clear theoretical framework to
demonstrate no presence of the geometric potential for the Dirac fermions.

For a particle that is constrained to remain on an ($N-1$)-dimensional curved
surface $\Sigma^{N-1}$ in the flat space $R^{N}$ ($N=2,3,4,...$), whether the
momentum and the kinetic energy operators must be extended to include the
possible contribution of curvatures has been controversial for quite a long
time. Part of the problem is the not unique form of the Hamiltonian after
quantization, and for a review see \cite{Golovnev}. Take a non-relativistic
particle on a surface for instance, Ikegami, Nagaoka, Takagi, and Tanzawa in
1992 showed that a quantum potential\ must certainly be arisen \cite{ikegami},
but Kleinert and Shabanov in 1997 demonstrated that no additional potential is
permissible from the so-called proper Dirac quantization of a free particle on
an ($N-1$)-dimensional sphere $S^{N-1}$ in $R^{N}$ \cite{Kleinert}. However,
since 2000, the disputes over such a \textit{geometric potential} exists for
the non-relativistic motion on curved hypersurface have gradually diminished.
Especially, during 2010\ to 2015, the physical consequences resulting from
extrinsic-curvature-dependent \textit{geometric potential }%
\cite{jk,dacosta,fc}\textit{ }and\textit{ geometric momentum }%
\cite{liu07,liu11,liu13,liu132,liu133,liu134,liu15,liu18,liu19} are
experimentally confirmed \cite{exp1,exp2,exp}, respectively. Nevertheless, for
the constrained particle that moves relativistically, whether there is
curvature-induced quantum potential remains an open problem.
\cite{maraner,mat1992,BurgessJensen1993,mat1994,mat1997,mat2000,AtansovSaxena2011,Olpak2012,AtansovSaxena2015,jose}
There is a mini-review on this subject available in a recent paper
\cite{liu19}.

Our principle to explore this problem is simple if not the simplest: \emph{All
symmetries expressed by the Poisson or Dirac brackets in classical mechanics
preserve in quantum mechanics; and so the Hamiltonian itself is also
determined by the symmetries. }\cite{liu11,liu19} This is an enlargement of
the Dirac quantization scheme. Let us first see what the usual Fundamental
quantum conditions (FQCs) are for a particle that moves in flat space $R^{N}$.
In this simplest case, our principle is the conventional Dirac quantization
scheme in which FQCs $[x_{i},x_{j}]=0,$ $[x_{i},p_{j}]=i\hbar\delta_{ij},$ and
$[p_{i},p_{j}]=0$ suffice, which are defined by the commutation relations
between positions $x_{i}$ and momenta $p_{i}$ ($i,j,k,l=1,2,3,...,N$) where
$N$ denotes the number of dimensions of the flat space in which the particle
moves\ \cite{Dirac}. In position representation, the momentum operator takes
simple form as $\mathbf{p}=-i\hslash\nabla$ where $\nabla\equiv\mathbf{e}%
_{i}\partial/\partial x_{i}$ is the ordinary gradient operator, and $N$
mutually orthogonal unit vectors $\mathbf{e}_{i}$ span the $N$ dimensional
Euclidean space $R^{N}$. Hereafter the Einstein summation convention over
repeated indices is used. Once the particle is constrained to remain on a
hypersurface $\Sigma^{N-1}$ embedded in $R^{N}$, the FQCs
become\ \cite{weinberg},%
\begin{equation}
\lbrack x_{i},x_{j}]=0,\text{ }[x_{i},p_{j}]=i\hbar(\delta_{ij}-n_{i}%
n_{j}),\text{ and }[p_{i},p_{j}]=-i\hbar\left\{  ({n_{i}}{n_{k}}_{,j}-{n_{j}%
}{n_{k}}_{,i})p_{k}\right\}  _{Hermitian}, \label{fqc}%
\end{equation}
where $O_{_{Hermitian}}$ stands for a Hermitian operator of an observable $O$,
and the equation of surface $f(\mathbf{x})=0$ can be so chosen that
$\left\vert \nabla f(\mathbf{x})\right\vert =1$ so $\mathbf{n\equiv}\nabla
f(\mathbf{x})=$ $\mathbf{e}_{i}n_{i}$ being the normal at a local point on the
surface. FQCs (\ref{fqc}) can by no mean give the unique form of the momentum
operators, and thus the construction of the unambiguous form of the
Hamiltonian operator is certainly impossible within the FQCs \cite{Golovnev}.
It is then reasonable to introduce more quantum conditions that together with
the FQCs must be utilized as first principles. Remember that in classical
mechanics, the classical brackets between $(\mathbf{x},H)_{cb}$ and
$(\mathbf{p},H)_{cb}$ can be easily computed, where the subscripts "$cb"$ mean
classical brackets, Poisson and Dirac brackets for instance. In quantum
mechanics, commutation relations $[\mathbf{x},H]=i\hbar\left\{  (\mathbf{x}%
,H)_{cb}\right\}  _{Hermitian}$ and $\left[  \mathbf{p},H\right]
=-i\hbar\left\{  (\mathbf{p},H)_{cb}\right\}  _{Hermitian}$ or \emph{their
derived relations} without operator-ordering problem (\textit{c.f.}
(\ref{dqc})) are hypothesized to be requirements upon the form of the
Hamiltonian operator $H$. Our principle for the constrained particle that
moves non-relativistically leads to the curvature-induced geometric potential
\cite{liu18}. The first application of the principle to relativistic motion is
for a particle that is on $2$D surface of sphere \cite{liu19}. The present
paper is the second application to the relativistic motion.

Once the motion is relativistically, we have following Dirac brackets\emph{
}containing classical brackets between $(\mathbf{x},H)$ and $(\mathbf{p},H)$
in the following, $p_{i}=H\left[  x_{i},H\right]  _{D}/c^{2}$,
and\ $\mathbf{n}\wedge\left[  \mathbf{p},H\right]  _{D}=0$ \cite{liu19,liu16},
where $\left[  f,g\right]  _{D}$ denotes Dirac bracket for two classical
quantities $f$ and $g$. The meaning of $\mathbf{n}\wedge\left[  \mathbf{p}%
,H\right]  _{D}=0$ is simple: The free particle on the surface experiences no
tangential force. While performing quantization, we have the so-called
dynamical quantum conditions (DQCs) \cite{liu19} accordingly,%
\begin{equation}
p_{i}=\frac{1}{i\hbar}\frac{H\left[  x_{i},H\right]  +\left[  x_{i},H\right]
H}{2c^{2}},and\ \mathbf{n}\wedge\left[  \mathbf{p},H\right]  -\left[
\mathbf{p},H\right]  \wedge\mathbf{n}=0. \label{dqc}%
\end{equation}
This set of DQCs imposes restrictions on the form of the Hamiltonian operator.
The FQCs and DQCs are the manifestation of our principle for the particle
moves relativistically on a hypersurface. To note that the form of generally
covariant momentum applicable to the spin particle is easily attainable with a
simple inclusion of the spin-connection contribution into the geometric
momentum that is originally applicable to the spinless particle \cite{liu19}.
The generally covariant geometric momentum is in general \cite{liu19}
\begin{equation}
{\mathbf{p}}=-i\hbar({\nabla_{\Sigma}}+\frac{{M{\mathbf{n}}}}{2}%
+i\mathbf{x}^{\mu}\Omega_{\mu})=-i\hbar({\nabla_{\Sigma}}+\frac{{M{\mathbf{n}%
}}}{2})+\hbar\mathbf{x}^{\mu}\Omega_{\mu}, \label{GM}%
\end{equation}
where $\mathbf{x}^{\mu}\equiv\partial\mathbf{x}/\partial\xi^{\mu}$ with
$\xi^{\mu}=(\xi^{1},\xi^{2},...,\xi^{N-1})$ being local parameters of the
surface $f(\mathbf{x})=0$ and $\mathbf{x=x}(\xi^{1},\xi^{2},...,\xi^{N-1})$,
$\Omega_{\mu}=\left(  -i/8\right)  \omega_{\mu}^{ab}\left[  \gamma_{a}%
,\gamma_{b}\right]  $ in which $\omega_{\mu}^{ab}$ are the spin-connections
\cite{Iorio,RMP1957,ogawa,Abrikosov} and $\gamma_{a}$ ($a,b=0,1,2,...N$) are
Dirac spin matrices, and ${\nabla_{\Sigma}\equiv}\mathbf{e}_{i}(\delta
_{ij}-n_{i}n_{j})\partial_{j}$ $=\nabla-\mathbf{n}\partial_{n}$ $=\mathbf{x}%
^{\mu}\partial_{\mu}$ is the the gradient operator, and the mean curvature
${M\equiv-\nabla_{\Sigma}}\cdot\mathbf{n}$ is defined by the sum of the all
principal curvatures. Without the spin-connection term, ${\mathbf{p}}$
(\ref{GM}) reduces to be $-i\hbar({\nabla_{\Sigma}}+{M{\mathbf{n/}}}2)$
\cite{liu07,liu11,liu13,liu132,liu18,exp,gem16,gem17,gem18}. The rest problem
is then to determine the general form of the quantum potential from DQCs
(\ref{dqc}). Unfortunately it turns out to be a formidable task for we
encounter great computational difficulties. However, for a Dirac fermion on
a\ curved surface of revolution $\Sigma^{2}$ in the flat space $R^{3}$, the
calculations are straightforward, and we can show that no quantum potential is admissible.

This paper is organized as follows. In section II, we are going to show that
how to apply both FQCs (\ref{fqc}) and DQCs (\ref{dqc}) for the Dirac fermion
on a\ curved surface of revolution $\Sigma^{2}$, resulting in no geometric
potential. In Section III we conclude the present study.

\section{A Dirac fermion on a\ curved surface of revolution}

The curved surface of revolution is with $u\in R$, $v\in\left[  0,2\pi\right)
$%
\begin{equation}
x=u\cos v;\text{ }y=u\sin v;\text{ }z=f(u). \label{222}%
\end{equation}
The metric tensor and the natural diagonal zweibein on a\ curved surface of
revolution are%
\begin{equation}
g_{\mu\nu}=diag\left(  1+f^{\prime2}(u),u^{2}\right)  ,\text{ }e_{\mu}%
^{a}=diag\left(  \sqrt{1+f^{\prime2}(u)},u\right)  . \label{223}%
\end{equation}
The nonzero components of spin connection are%
\begin{equation}
\omega_{v}^{12}=-\omega_{v}^{21}=-\frac{1}{\sqrt{1+f^{\prime2}(u)}}.
\label{224}%
\end{equation}
The generally covariant derivatives are then%
\begin{equation}
{{\nabla}_{u}}={{\partial}_{u}}\ \ \ and\ \ \ {{\nabla}_{v}}={{\partial}_{v}%
}-\frac{i}{2}\frac{{{\sigma}_{z}}}{\sqrt{1+f^{\prime2}(u)}}. \label{225}%
\end{equation}
In final, the relativistic Hamiltonian operator $H_{0}=i\hbar\gamma
^{a}{{\nabla}_{a}=}i\hbar\gamma^{a}e_{a}^{\mu}{{\nabla}_{\mu}}$ \cite{RMP1957}
without geometric potential becomes%
\begin{equation}
H_{0}=-i\hbar({{\sigma}_{x}}(\frac{1}{\sqrt{1+f^{\prime2}(u)}}{{\partial}%
_{u}+}\frac{1}{2u\sqrt{1+f^{\prime2}(u)}})+{{\sigma}_{y}}\frac{1}{u}%
{{\partial}_{v}).} \label{226}%
\end{equation}
In quantum mechanics, the general form of the Hamiltonian must be assumed to
be ${H}=H_{0}+{{V}_{G}}$ where ${{V}_{G}}$ will be discussed shortly.

The generally covariant geometric momenta (\ref{GM}) now give,%
\begin{subequations}
\begin{align}
{{p}_{x}}  &  =-i\hbar\left(  \frac{\cos v}{1+f^{\prime2}(u)}{{\partial}_{u}%
}-\frac{\sin v}{u}{{\partial}_{v}}-\frac{\cos v(f^{\prime2}(u)+f^{\prime
4}(u)+uf^{\prime}(u)f^{\prime\prime}(u))}{2u\left(  1+f^{\prime2}(u)\right)
^{2}}\right)  +\frac{\hbar\sin v}{u}\frac{{{\sigma}_{z}}}{2\sqrt{1+f^{\prime
2}(u)}},\label{gmx}\\
{p_{y}}  &  =-i\hbar\left(  \frac{\sin v}{1+f^{\prime2}(u)}{{\partial}_{u}%
}+\frac{\cos v}{u}{{\partial}_{v}}-\frac{\sin v(f^{\prime2}(u)+f^{\prime
4}(u)+uf^{\prime}(u)f^{\prime\prime}(u))}{2u\left(  1+f^{\prime2}(u)\right)
^{2}}\right)  -\frac{\hbar\cos v}{u}\frac{{{\sigma}_{z}}}{2\sqrt{1+f^{\prime
2}(u)}},\label{gmy}\\
{p}_{z}  &  =-i\hbar\left(  \frac{f^{\prime}(u)}{1+f^{\prime2}(u)}{{\partial
}_{u}{+}}\frac{f^{\prime}(u)+f^{\prime3}(u)+uf^{\prime\prime}(u)}{2u\left(
1+f^{\prime2}(u)\right)  ^{2}}\right)  . \label{gmz}%
\end{align}
The spin-connections are equivalent to a gauge potential $\mathbf{A}$\ whose
components are explicitly,%
\end{subequations}
\begin{equation}
{{A}_{x}}=-\frac{\hbar\sin v}{u}\frac{{{\sigma}_{z}}}{2\sqrt{1+f^{\prime2}%
(u)}},{{A}_{y}}=\frac{\hbar\cos v}{u}\frac{{{\sigma}_{z}}}{2\sqrt
{1+f^{\prime2}(u)}},{{A}_{z}}=0. \label{227}%
\end{equation}
It is compatible with previous results on relationship between
spin-connections for fermions on curved surface and gauge fields
\cite{2010PR,LeeDH,Iorio,RMP1957,ogawa}.

Considering the orthogonality and completeness of the $2\ast2$
matrices\ ($I,{{\sigma}_{x}},{{\sigma}_{y}},{{\sigma}_{z}}$), we can assume
that the geometric potential ${{V}_{G}}$ takes the following most general
form,%
\begin{equation}
{{V}_{G}}={{a}_{0}}I+{{a}_{x}}{{\sigma}_{x}}+{{a}_{y}}{{\sigma}_{y}}+{{a}_{z}%
}{{\sigma}_{z},} \label{228}%
\end{equation}
where $\left(  {{a}_{0}},\ {{a}_{x}},\ {{a}_{y}},\ {{a}_{z}}\right)  $ are
ansatz functions of $u$ and $v$ to be determined via requirements (\ref{dqc}).
Three commutators $\left[  {{p}_{i}},{H}\right]  $ and the results are,
respectively,%
\begin{subequations}
\begin{align}
\left[  {{p}_{x}},{H}\right]   &  =\frac{{{\hbar}^{2}}}{2u^{2}(1+f^{\prime
2}(u))^{4}}(-{{\sigma}_{y}}f^{\prime2}(u)\left(  1+3f^{\prime2}(u)+3f^{\prime
4}(u)+f^{\prime6}(u)\right)  \left(  2\cos v\partial_{v}-\sin v\right)
\nonumber\\
&  \text{ \ \ \ \ \ \ \ \ \ \ \ \ \ \ \ \ \ \ \ \ \ \ }-{{\sigma}_{x}}u\cos
vf^{\prime}(u)\left(  1+f^{\prime2}(u)\right)  ^{\frac{3}{2}}\left(
f^{\prime\prime}(u)+uf^{3}\left(  u\right)  +2uf^{\prime\prime}(u\right)
\partial_{u})\nonumber\\
&  \text{ \ \ \ \ \ \ \ \ \ \ \ \ \ \ \ \ \ \ \ \ \ \ }+{{\sigma}_{x}}%
u^{2}\cos v\left(  1+f^{\prime2}(u)\right)  ^{\frac{1}{2}}f^{\prime\prime
2}(u)\left(  3f^{\prime2}(u)-1\right)  )+\left[  {{p}_{x}},{{V}_{G}}\right]
,\label{pxh}\\
\left[  {{p}_{y}},{H}\right]   &  =\frac{{{\hbar}^{2}}}{2u^{2}(1+f^{\prime
2}(u))^{4}}(-{{\sigma}_{y}}f^{\prime2}(u)\left(  1+3f^{\prime2}(u)+3f^{\prime
4}(u)+f^{\prime6}(u)\right)  \left(  2\sin v\partial_{v}+\cos v\right)
\nonumber\\
&  \text{ \ \ \ \ \ \ \ \ \ \ \ \ \ \ \ \ \ \ \ \ \ \ }-{{\sigma}_{x}}u\sin
vf^{\prime}(u)\left(  1+f^{\prime2}(u)\right)  ^{\frac{3}{2}}\left(
f^{\prime\prime}(u)+uf^{3}\left(  u\right)  +2uf^{\prime\prime}(u\right)
\partial_{u})\nonumber\\
&  \text{ \ \ \ \ \ \ \ \ \ \ \ \ \ \ \ \ \ \ \ \ \ \ }+{{\sigma}_{x}}%
u^{2}\sin v\left(  1+f^{\prime2}(u)\right)  ^{\frac{1}{2}}f^{\prime\prime
2}(u)\left(  3f^{\prime2}(u)-1\right)  )+\left[  {{p}_{y}},{{V}_{G}}\right]
,\label{pyh}\\
\left[  {{p}_{z}},{H}\right]   &  =\frac{{{\hbar}^{2}}}{2u^{2}(1+f^{\prime
2}(u))^{\frac{7}{2}}}\left(  {{\sigma}_{x}}u\left(  uf^{\prime\prime
}(u\right)  \partial_{u}+(1+f^{\prime2}(u))\left(  f^{\prime\prime}%
(u)+uf^{3}\left(  u\right)  \right)  -4uf^{\prime}(u)f^{\prime\prime
2}(u)\right) \nonumber\\
&  \text{ \ \ \ \ \ \ \ \ \ \ \ \ \ \ \ \ \ \ \ \ \ \ \ }+{{\sigma}_{y}%
}f^{\prime}(u)\left(  1+f^{\prime2}(u)\right)  ^{\frac{3}{2}}\partial
_{v})+\left[  {{p}_{z}},{{V}_{G}}\right]  . \label{333}%
\end{align}
During the calculations, we find that$\ \mathbf{n}\wedge\left[  \mathbf{p}%
,H_{0}\right]  -\left[  \mathbf{p},H_{0}\right]  \wedge\mathbf{n}=0$. It
strongly implies that no geometric potential is necessarily introduced. To see
it, let us first compute the following commutation relations $\left[  {{p}%
_{i}},{{V}_{G}}\right]  $,%
\end{subequations}
\begin{subequations}
\begin{align}
\left[  {{p}_{x}},{{V}_{G}}\right]   &  =-i\hbar\left(  \frac{\cos
v}{1+f^{\prime2}(u)}{{\partial}_{u}{V}_{G}}-\frac{\sin v}{u}{{\partial}_{v}%
{V}_{G}}\right)  +i\hbar\frac{\sin v}{u}\frac{{{1}}}{\sqrt{1+f^{\prime2}(u)}%
}\left(  {{a}_{x}}{{\sigma}_{y}}-{{a}_{y}}{{\sigma}_{x}}\right)
,\label{pxv}\\
\lbrack{{p}_{y}},{{V}_{G}]}  &  =-i\hbar\left(  \frac{\sin v}{1+f^{\prime
2}(u)}{{\partial}_{u}{V}_{G}}+\frac{\cos v}{u}{{\partial}_{v}{V}_{G}}\right)
-i\hbar\frac{\cos v}{u}\frac{{{1}}}{\sqrt{1+f^{\prime2}(u)}}\left(  {{a}_{x}%
}{{\sigma}_{y}}-{{a}_{y}}{{\sigma}_{x}}\right)  ,\label{pyv}\\
\left[  {{p}_{z}},{{V}_{G}}\right]   &  =-i\hbar\frac{f^{\prime}%
(u)}{1+f^{\prime2}(u)}{{\partial}_{u}{V}_{G}.} \label{229}%
\end{align}

Three components of the vector equations $\mathbf{n}\wedge\left[
\mathbf{p},H\right]  -\left[  \mathbf{p},H\right]  \wedge\mathbf{n=}0\ $reduce
to $\mathbf{n}\wedge\left[  \mathbf{p},\ {{V}_{G}}\right]  -\left[
\mathbf{p},\ {{V}_{G}}\right]  \wedge\mathbf{n=}0$. Explicitly, we have,%
\end{subequations}
\begin{subequations}
\begin{align}
&  2i\hbar\left(  \frac{\sin v}{\sqrt{1+f^{\prime2}(u)}}{{\partial}_{u}{V}%
_{G}}+\frac{\cos v}{u\sqrt{1+f^{\prime2}(u)}}{{\partial}_{v}}{{V}_{G}}%
+\frac{\cos v}{u}\frac{{{1}}}{\left(  1+f^{\prime2}(u)\right)  }\left(
{{a}_{x}}{{\sigma}_{y}}-{{a}_{y}}{{\sigma}_{x}}\right)  \right)  =0,\\
&  2i\hbar\left(  -\frac{\cos v}{\sqrt{1+f^{\prime2}(u)}}{{\partial}_{u}%
{V}_{G}}+\frac{\sin v}{u\sqrt{1+f^{\prime2}(u)}}{{\partial}_{v}}{{V}_{G}%
}+\frac{\sin v}{u}\frac{{{1}}}{\left(  1+f^{\prime2}(u)\right)  }\left(
{{a}_{x}}{{\sigma}_{y}}-{{a}_{y}}{{\sigma}_{x}}\right)  \right)  =0,\\
&  2i\hbar\left(  \frac{f^{\prime}(u)}{u\sqrt{1+f^{\prime2}(u)}}{{\partial
}_{v}}{{V}_{G}}+\frac{f^{\prime}(u)}{u\left(  1+f^{\prime2}(u)\right)
}\left(  {{a}_{x}}{{\sigma}_{y}}-{{a}_{y}}{{\sigma}_{x}}\right)  \right)  =0.
\label{330}%
\end{align}
After simplification, we have,
\end{subequations}
\begin{subequations}
\begin{align}
&  \left(  \sin v{{\partial}_{u}}+\frac{\cos v}{u}{{\partial}_{v}}\right)
{{V}_{G}}+\frac{\cos v}{u\sqrt{1+f^{\prime2}(u)}}\left(  {{a}_{x}}{{\sigma
}_{y}}-{{a}_{y}}{{\sigma}_{x}}\right)  =0\\
&  \left(  -\cos v{{\partial}_{u}}+\frac{\sin v}{u}{{\partial}_{v}}\right)
{{V}_{G}}+\frac{\sin v}{u\sqrt{1+f^{\prime2}(u)}}\left(  {{a}_{x}}{{\sigma
}_{y}}-{{a}_{y}}{{\sigma}_{x}}\right)  =0\\
&  {{\partial}_{v}}{{V}_{G}}+\frac{1}{\sqrt{1+f^{\prime2}(u)}}\left(  {{a}%
_{x}}{{\sigma}_{y}}-{{a}_{y}}{{\sigma}_{x}}\right)  =0
\end{align}
The general solutions are $\left(  {{a}_{0}},\ {{a}_{x}},\ {{a}_{y}}%
,\ {{a}_{z}}\right)  =\left(  c_{1},0,0,c_{2}\right)  $ where $c_{1}$ and
$c_{2}$ are two constants, i.e., ${{V}_{G}}=c_{1}+c_{2}{{\sigma}_{z}%
=}diag\left(  c_{1}+c_{2},c_{1}-c_{2}\right)  $. These constants can be set as
zero as it is matter of shifting the reference point of the energy. In other
words, there is no geometric potential.

Three examples are in the following.

Example one: A Dirac fermion on Torus. The toroidal surface is with two local
coordinates $\theta\in\left[  0,2\pi\right)  ,\varphi\in\left[  0,2\pi\right)
$,%
\end{subequations}
\begin{equation}
x=\left(  R+r\sin\theta\right)  \cos\varphi;\text{ }y=\left(  R+r\sin
\theta\right)  \sin\varphi;\text{ }z=r\cos\theta,\left(  R>r\neq0\right)  ,
\label{16}%
\end{equation}
where $\varphi$ is the azimuthal angle and $\theta$ the polar angle, and $R$
and $r$ are the outer and inner radii of the torus, respectively. Three
equations for geometric potential ${{V}_{G}}$ are, respectively,
\begin{subequations}
\begin{align}
\left(  \frac{\sin\varphi}{r}{{\partial}_{\theta}}+\frac{{{\cos}}\theta
\cos\varphi}{R+r\sin\theta}{{\partial}_{\varphi}}\right)  {{V}_{G}}%
+\frac{{{\cos}^{2}}\theta\cos\varphi}{R+r\sin\theta}\left(  {{a}_{x}}{{\sigma
}_{y}}-{{a}_{y}}{{\sigma}_{x}}\right)   &  =0\\
\left(  -\frac{\cos\varphi}{r}{{\partial}_{\theta}}+\frac{{{\cos}}\theta
\sin\varphi}{R+r\sin\theta}{{\partial}_{\varphi}}\right)  {{V}_{G}}%
+\frac{{{\cos}^{2}}\theta\sin\varphi}{R+r\sin\theta}\left(  {{a}_{x}}{{\sigma
}_{y}}-{{a}_{y}}{{\sigma}_{x}}\right)   &  =0\\
{{\partial}_{\varphi}}{{V}_{G}}+\cos\theta\left(  {{a}_{x}}{{\sigma}_{y}}%
-{{a}_{y}}{{\sigma}_{x}}\right)   &  =0
\end{align}
The orthogonality and completeness of the $2\ast2$ matrices\ ($I,{{\sigma}%
_{x}},{{\sigma}_{y}},{{\sigma}_{z}}$) imply that there is no geometric
potential, i.e., ${{V}_{G}}=0$.

Example two: A Dirac fermion on Catenoid. The catenoid is with two local
coordinates $\theta\in\left[  0,\ 2\pi\right)  ,\ \rho\in R$,%
\end{subequations}
\begin{equation}
x=a\cosh\frac{\rho}{a}\cos\theta;\ y=a\cosh\frac{\rho}{a}\sin\theta
;\ z=\rho,\ \left(  a>0\right)  \label{28}%
\end{equation}
where $a$ is the constant. Three equations for geometric potential ${{V}_{G}}$
are, respectively,%
\begin{subequations}
\begin{align}
\left(  \tanh\frac{\rho}{a}{{\partial}_{\theta}}+a\tan\theta{{\partial}_{\rho
}}\right)  {{V}_{G}}-{{\tanh}^{2}}\frac{\rho}{a}\left(  {{a}_{x}}{{\sigma}%
_{y}}-{{a}_{y}}{{\sigma}_{x}}\right)   &  =0\\
\left(  \tanh\frac{\rho}{a}{{\partial}_{\theta}}-a\cot\theta{{\partial}_{\rho
}}\right)  {{V}_{G}}-{{\tanh}^{2}}\frac{\rho}{a}\left(  {{a}_{x}}{{\sigma}%
_{y}}-{{a}_{y}}{{\sigma}_{x}}\right)   &  =0\\
{{\partial}_{\theta}}{{V}_{G}}-\tanh\frac{\rho}{a}\left(  {{a}_{x}}{{\sigma
}_{y}}-{{a}_{y}}{{\sigma}_{x}}\right)   &  =0
\end{align}
Again, there is no geometric potential, i.e., ${{V}_{G}}=0$.

Example three: A fermion on the symmetric ellipsoid. The symmetric ellipsoid
is with two local coordinates $\theta\in\left(  0,\pi\right)  ,\varphi
\in\left(  0,2\pi\right)  $,%
\end{subequations}
\[
x=a\sin\theta\cos\varphi;y=a\sin\theta\sin\varphi;z=c\cos\theta,
\]
where $a$ and $c$ are constant. Three equations for geometric potential
$V_{g}$ are, respectively,%
\begin{subequations}
\begin{align}
\left(  \sin\varphi\partial_{\theta}+\frac{\cos\theta\cos\varphi}{\sin\theta
}\partial_{\varphi}\right)  V_{g}+\frac{a\cos^{2}\theta\cos\varphi}{\sin
\theta\sqrt{\left(  a\cos\theta\right)  ^{2}+\left(  c\sin\theta\right)  ^{2}%
}}\left(  {{a}_{x}}{{\sigma}_{y}}-{{a}_{y}}{{\sigma}_{x}}\right)   &  =0\\
\left(  -\cos\varphi\partial_{\theta}+\frac{\cos\theta\sin\varphi}{\sin\theta
}\partial_{\varphi}\right)  V_{g}+\frac{a\cos^{2}\theta\sin\varphi}{\sin
\theta\sqrt{\left(  a\cos\theta\right)  ^{2}+\left(  c\sin\theta\right)  ^{2}%
}}\left(  {{a}_{x}}{{\sigma}_{y}}-{{a}_{y}}{{\sigma}_{x}}\right)   &  =0\\
\frac{c}{a}\partial_{\varphi}V_{g}+\frac{c\cos\theta}{\sqrt{\left(
a\cos\theta\right)  ^{2}+\left(  c\sin\theta\right)  ^{2}}}\left(  {{a}_{x}%
}{{\sigma}_{y}}-{{a}_{y}}{{\sigma}_{x}}\right)   &  =0
\end{align}
We see ${{V}_{G}}=0$ as well.

\section{Discussions and Conclusions}

Surface quantum states exist in many systems. Surface plasmon polaritons and
topological insulators are two typical ones. So, we must understand the
behaviors of free particle constrained on an $\left(  N-1\right)
$-dimensional curved surface $\Sigma^{N-1}$ embedded in $N$-dimensional flat
space $R^{N}$. Once the (quasi-)particle moves non-relativistically, we have
the geometric potential. Once it moves relativistically instead, there may be
no such geometric potential. The present study explicitly shows that an
enlargement of the Dirac quantization scheme can give definite results that
for the relativistic spin $1/2$ particles constrained on the two-dimensional
curved surfaces of revolution, no geometric potential exists.

There are three problems that remain open. One is that usually we take it for
granted that non-relativistic motion in flat space is the limit of the
relativistic motion in it. Once the motion is constrained, the relation in
between is not clear. The second is that in present paper we deal with spin
$1/2$ particle, what about an arbitrary boson or fermion under constrained and
relativistic motion has been unknown yet. The third is: we are confident that
Dirac quantization scheme is sufficient, but the number of quantum conditions
must be finite. Though FQCs and DQCs are sufficient for non-relativistic
particles in general and relativistic spin $1/2$ particles for some types of
surfaces as shown in present paper, we do not know whether they are sufficient
for the relativistic particles constrained on $\Sigma^{N-1}$.
\end{subequations}
\begin{acknowledgments}
This work is financially supported by National Natural Science Foundation of
China under Grant No. 11675051.
\end{acknowledgments}

\end{document}